\documentclass[prl,showpacs,amsmath,twocolumn,superscriptaddress,letterpaper,floatfix]{revtex4} 
\usepackage{amssymb,verbatim}
\usepackage{amsmath}
\usepackage{graphicx}
\usepackage[colorlinks]{hyperref} %links for citations
\usepackage{mathptm,times} 
\usepackage{units}
\usepackage{bbm}
\newcommand{\tr}{\mbox{tr}}
\newcommand{\ket}[1]{\left | #1 \right \rangle}
\newcommand{\bra}[1]{\left \langle #1 \right |}

\newcommand{\beq}{\begin{equation}}
\newcommand{\eeq}{\end{equation}}
\newcommand{\beqa}{\begin{eqnarray}}
\newcommand{\eeqa}{\end{eqnarray}}

\begin{document}
\title{Statistical dynamics of a non-Abelian anyonic quantum walk}
\author{Lauri Lehman}
\affiliation{Centre for Quantum Information Science and Security, Macquarie
University, 2109, NSW Australia}
\affiliation{School of Physics and Astronomy, University of Leeds, Leeds LS2 9JT, UK}
\author{Vaclav Zatloukal}
\affiliation{School of Physics and Astronomy, University of Leeds, Leeds LS2 9JT, UK}
\author{Gavin K. Brennen}
\affiliation{Centre for Quantum Information Science and Security, Macquarie
University, 2109, NSW Australia}
\affiliation{School of Physics and Astronomy, University of Leeds, Leeds LS2 9JT, UK}
\author{Jiannis K. Pachos}
\affiliation{School of Physics and Astronomy, University of Leeds, Leeds LS2 9JT, UK}
\author{Zhenghan Wang}
\affiliation{Microsoft Research, Station Q, University of California,
Santa Barbara, CA 93106, USA}

%\affiliation{Centre for Quantum Information Science and Security, Macquarie
%University, 2109, NSW Australia}
%\affiliation{School of Physics and Astronomy, University of Leeds, Leeds LS2 9JT, UK}

\date{\today}  

\begin{abstract}
We study the single particle dynamics of a mobile non-Abelian anyon hopping around many pinned anyons on a surface.  The dynamics is modelled by a discrete time quantum walk and the spatial degree of freedom of the mobile anyon becomes entangled with the fusion degrees of freedom of the collective system.  
Each quantum trajectory makes a closed braid on the world lines of the particles establishing a direct connection between statistical dynamics and quantum link invariants.  
We find that asymptotically a mobile Ising anyon becomes so entangled with its environment that its statistical dynamics reduces to a classical random walk with linear dispersion in contrast to particles with Abelian statistics which have quadratic dispersion.    

\end{abstract}

\maketitle
%%%%%%%%%%%%%%%%%%%%%%%%%%%%%%%%%%%%%%%%%%%%%%%%%
%{\it Introduction.---}

Anyons are point like particles with more general statistics than bosons or fermions.  They
were shown to exist in systems where the physics is constrained
to two dimensions \cite{Leinass}.    Beyond mere possible existence they where found to 
be a good description for low lying quasi-particle excitations of fractional quantum Hall systems \cite{Arovas, Moore} and they exactly describe excitations in various strongly correlated two dimensional spin lattice models \cite{Wen, Kitaev}. 
Recently there has been tremendous experimental progress in preparation and control of systems capable of exhibiting topological order \cite{Dolev, Bloch} with the goal to observe anyonic statistics.  This is further motivated by the discovery that braiding some types of non-Abelian anyons can be used for naturally fault tolerant quantum computing \cite{Freedman}.  The quantum physics of anyonic systems is very rich but is only beginning to be explored in its own right.  For example, there have been investigations of the equilibrium properties of dynamically interacting, but static, non-Abelian anyons in chains  \cite{Feiguin} and two dimensional lattices \cite{Ludwig, Pachos}.   

Here we describe a simple model which captures some of the non-equilibrium physics of moving non-Abelian anyons interacting purely due to particle statistics.   In Ref. \cite{BEKPTW} the authors introduced a general protocol for quantum walks with anyons to describe the dynamics of one mobile anyon braiding about other pinned anyons on a surface.  It was shown that while for Abelian
anyons the dispersion is quadratic as in the usual quantum walk, the non-Abelian walk appears to have richer behavior.  Transition from coherent quantum to classical random behavior with linear dispersion occurs when a quantum walk strongly decoheres due to interaction with an environment \cite{Kendon}. Indeed, Ref. \cite{Brun} found that if one introduces a new coin at every other step, or fewer, in a quantum walk the dispersion is quadratic but if a new coin is introduced every time step  then the dispersion is linear.  One might expect that statistical interactions of non-Abelian anyons would be sufficient to induce such a transition.  We show that this is the case 
by expressing the statistical dynamics of the mobile anyon as a function of topological invariants of the links of the anyonic worldlines generated during the quantum walk.  Specifically, we solve for the asymptotic distribution of the Ising model non-Abelian anyons, $\sigma$, which appear as quasi-particle excitations in the Pfaffian wave function description of the $5/2$ filled fractional quantum Hall state \cite{Moore,NayakC}.
It is the purpose of this work to determine the behaviour of such a system by analytical methods, thus opening the way for modelling complex systems that are of interest to statistical physics 
\cite{Nechaev}.  

\begin{figure}[h]
\begin{center}
\includegraphics[width=8cm]{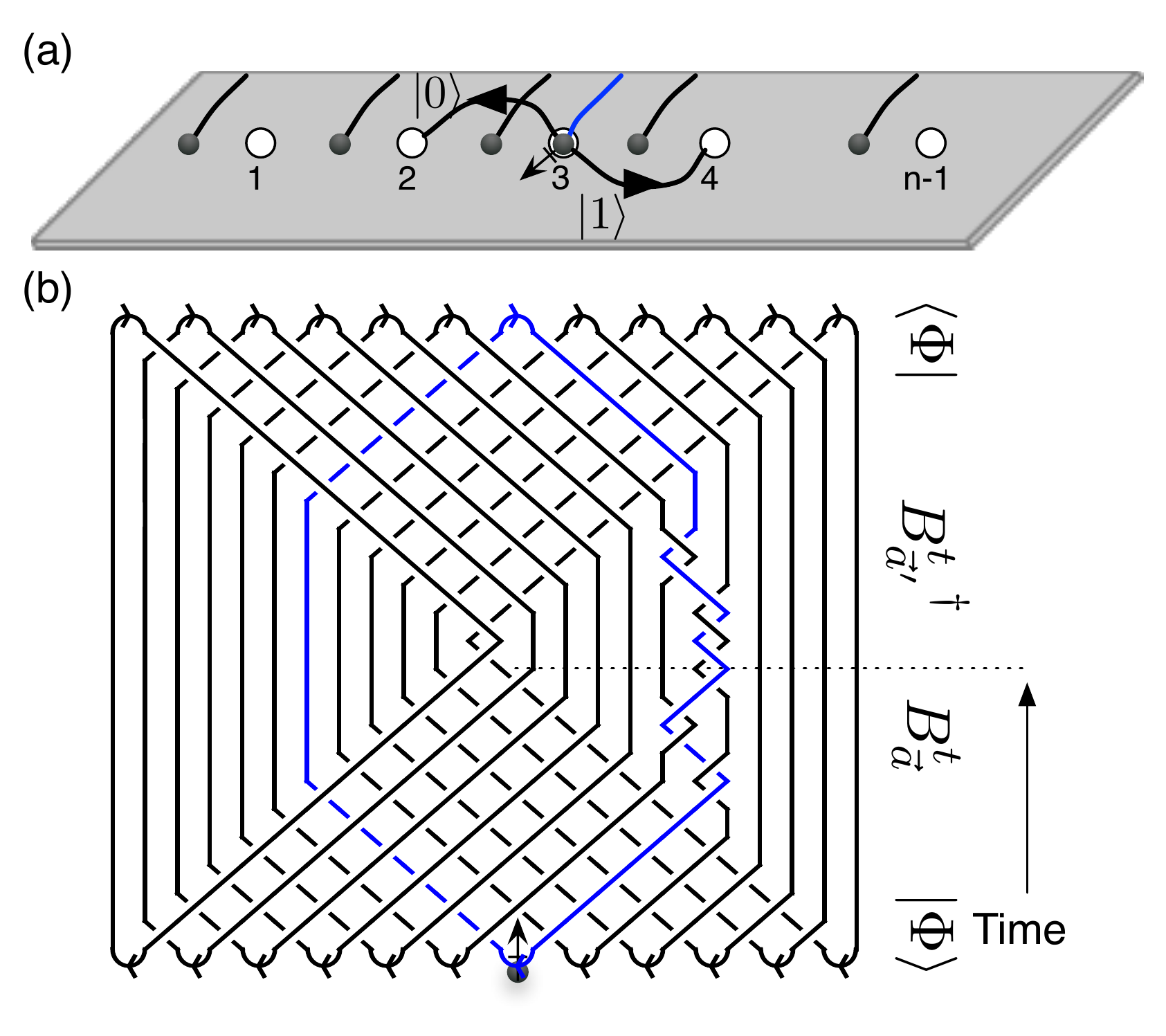}
\caption{\label{fig:1} Anyonic quantum walk in a system with $n$ vacuum pairs of anyons in state $\ket{\Phi}$.  (a) Half of each vacuum pair (depicted as connected by a string) participates with one walker anyon carrying a spin braiding around $n-1$ pinned anyons. (b)   Link representation of the world lines for a Markov closed quantum trajectory that contributes to the spatial distribution $p(3,5)$.  Here $\vec{a}=(1,0,0,1,1)^T$ and  $\vec{a}'=(0,1,1,0,1)^T$.  
%The contribution is proportional to the Kauffman bracket of the link $L$ which is the Markov trace of the braid word on the $n$ participating anyons or equivalently the expectation value of ${B_{\vec{a}'}^t}^\dagger B^t_{\vec{a}}$ in state $\ket{\Phi}$ of all $2n$ anyons.  
The link shown is proper and has one Borromean ring, i.e. if any of three linked components were cut the others would become disentangled.  Here $\mathtt{arf}(L)=-1$.}
\end{center}
\end{figure}

The setup (see Fig. \ref{fig:1}a) is a surface with $n$ vacuum pairs of anyons of topological charge $\sigma$, and one member of each pair participates in the dynamics.  The other $n$ anyons are moved out of the way or could be excitations on the boundary  \footnote{Here we assume the dynamics takes place on a $2$ sphere or a disk but equivalently it could be a surface with punctures and one mobile anyon}.  The participating anyons are canonically ordered on the surface with the $n-1$ pinned anyons and one mobile {\it walker} anyon.  The walker hops between neighboring sites with spatial index $s=1,\dots n-1$ and has an additional spin-$1/2$ degree of freedom (DOF) we dub a coin.   The pinned anyons are located in between the sites.  
 The total Hilbert space decomposes as $\mathcal{H}=\mathcal{H}_{\rm space}\otimes \mathcal{H}_{\rm coin}\otimes \mathcal{H}_{\rm fusion}\simeq\mathbb{C}^{n-1}\otimes\mathbb{C}^2\otimes \mathbb{C}^{D}$ which becomes infinite dimensional in the asymptotic limit.
%  The dimension of the topological fusion degrees of freedom scales exponentially with $n$ and is computed using the fusion rules for the anyons.
Fusion DOFs denumerate the number of distinct measurement outcomes of topological charge when pairs of anyons are fused together \cite{Freedman}.  Its size grows like $D\sim d^n$ where $d$ is the quantum dimension of the anyons.  For $n$ Ising anyons with total trivial charge, we have $d=\sqrt{2}$ and $D=2^{n/2-1}$, thus effectively they introduce half a new coin DOF per step of the walk.
%  made on pairs of anyons and are determined by the fusion rules $a_i\times
%a_j=\sum_k N_{a_ia_j}^{a_k}a_k$, where $N_{a_ra_s}^{a_t}\in~\mathbb{N}$ counts the
%number of ways to combine anyons of type $a_r$ and $a_s$ to obtain $a_t$.  For $n$ anyons of type $\sigma$ with trivial overall topological charge, the
%dimension of the fusion space is $D=\sum_{a_1,a_2,\ldots,a_{n-2}}N_{\sigma
%\sigma}^{a_1}N_{a_1\sigma}^{a_2}N_{a_2\sigma}^{a_3}\cdots
%N_{a_{n-2}\sigma}^{1}$ which scales like $d^n$ where $d$ is the quantum
%dimension of the $\sigma$ anyon.  

  The dynamics is modelled by a composition of two discrete unitary steps $W=TU$ where $U$ acts on the coin and $T$ is a conditional braiding operator.  It moves the walker to the right or left depending on the coin state:
\[
\begin{array}{lll}
T&=&\sum_{s=1}^{n-2} \ket{s-1}_{\rm space}\bra{s}\otimes \ket{0}_{\rm coin}\bra{0}\otimes b_{s-1}\\
&&+\ket{s+1}_{\rm space}\bra{s}\otimes\ket{1}_{\rm coin}\bra{1}\otimes b_{s}
\end{array}
\]
where $\{b_s\}_{s=1}^{n-1}$ is a set of unitary generators of the braid group $\mathcal{B}_n$.  The particular representation depends on the braiding and recoupling rules for the chosen anyons.
%, s is a unitary representation of the braid group generator that winds an anyon at position $s$ counterclockwise around the pinned anyon at position $\frac{2s+1}{2}$ or winds an anyon at position $s+1$ counterclockwise around the pinned anyon at position $\frac{2s-1}{2}$. 
Notice that the chirality of the mobile anyonic charge current is fixed counterclockwise by this walk.
To make $T$ unitary we assume periodic boundary conditions ($\ket{0}_{\rm space}\equiv\ket{n-1}_{\rm space}$) but will be concerned with walks satisfying $|n/2|<t$ so that winding around the surface is not an issue.

The system's initial state is $\ket{\Psi(0)}=\ket{s_0=\lceil \frac{n}{2}\rceil}_{\rm space}\ket{0}_{\rm coin}\ket{\Phi}_{\rm fusion}$ where $\ket{\Phi}$ is the vacuum configuration of the $n$ pairs of anyons with half the members braided to the right.   After $t$ iterations, the state is $\ket{\Psi(t)}=W^t \ket{\Psi(0)}$ and the reduced state of the spatial DOF of the walker is
% \begin{equation}
% \begin{array}{lll}
%\rho_{\rm space}(t)&=&\tr_{\rm coin} \tr_{\rm fusion} \ket{\Psi(t)}\bra{\Psi(t)}\\
%&=&\displaystyle{\sum_{\vec{a},\vec{a}'}
%\tr {\mathcal{U}}_{\vec{a}\vec{a}'}^t
%\tr \mathcal{Y}_{\vec{a}\vec{a}'}^t
%\big| 2|\vec{a}|-t+s_0\big\rangle\big\langle 2|\vec{a}'|-t+s_0\big|}
%\end{array}
%\label{walkerdis}
% \end{equation}

\begin{eqnarray}
\rho_{\rm space}(t)&&\!\!\!\!\!\!\!\!\!=\tr_{\rm coin} \tr_{\rm fusion} \ket{\Psi(t)}\bra{\Psi(t)}\nonumber\\
&&\!\!\!\!\!\!\!\!\!=l\displaystyle{\sum_{\vec{a},\vec{a}'}
\tr {\mathcal{U}}_{\vec{a}\vec{a}'}^t
\tr \mathcal{Y}_{\vec{a}\vec{a}'}^t
\big| 2|\vec{a}|-t+s_0\big\rangle\big\langle 2|\vec{a}'|-t+s_0\big|}
\label{walkerdis}
\end{eqnarray}
where
$
{\mathcal{U}}_{\vec{a}\vec{a}'}^t =(\prod_{r=1}^t P_{a_r}U) \ket{0}_{\rm coin}\bra{0}
(\prod_{r=1}^t P_{a'_r}U^{\dagger})
$
and
$
\mathcal{Y}_{\vec{a}\vec{a}'}^t =B_{\vec{a}}^t\ket{\Phi}\bra{\Phi}
{B_{\vec{a}'}^t}^\dagger.
$
where $\ket{\Phi}$ corresponds to the Markov trace state (see Fig. \ref{fig:1}b).
The coin histories are
given by the vectors $\vec{a},\vec{a}'\in
\{0,1\}^{\otimes t}$, and the projectors for each outcome are $P_{a_j}=\ket{a_j}_{\rm coin}\bra{a_j}$.  The braid word for a given coin history is
\begin{equation}
B^t_{\vec{a}}=\prod_{r=0}^{t-1}b_{s_0+a_{t-r}+2(\sum_{j=1}^{t-r-1}a_j)-(t-r)}.
\label{braidworda}
\end{equation}
For a
Hadamard coin flip operation $U=\frac{1}{\sqrt{2}}\left( \begin{smallmatrix} 1&1\\1&-1 \end{smallmatrix} \right)$, and
$
\tr {\mathcal{U}}_{\vec{a}\vec{a}'}^t = {1 \over 2^t} (-1)^{z(\vec{a},\vec{a}')}
$
where
$
z(\vec{a},\vec{a}')=\sum_{j=1}^{t-1}a'_ja'_{j+1}+a_ja_{j+1},
$
is the sum of pairs of consecutive right moves (or 1 outcomes of the coin). 
The trace over the fusion DOF can be related to the
Kauffman bracket of a link, denoted $\langle L
\rangle$ for a link $L$.  Here we get a link which is the
Markov trace over the braid words for the forward and backward time
evolution histories as dictated by Eq. \ref{walkerdis}:
\begin{equation}
L=({B_{\vec{a}'}^t}^\dagger B^t_{\vec{a}})^\text{Markov}.
\label{modfL}
\end{equation}
Moreover,
\[
\tr \mathcal{Y}_{\vec{a}\vec{a}'}^t =\frac{1}{d^{(n-1)}} \langle ({B_{\vec{a}'}^t}^\dagger
 B^t_{\vec{a}})^\text{Markov}\rangle
\]
where $d$ is the quantum dimension of the anyons \cite{BEKPTW}.
%, and 
%the braid word ${B_{\vec{a}'}^t}^\dagger B^t_{\vec{a}}$ has been Markov traced,
%which corresponds to evaluating the Kauffman bracket of the corresponding
%link of the world lines of the anyons.

Concerning ourselves with the diagonal elements of the spatial probability
distribution $p(s,t)\equiv \bra{2s-t+s_0}\rho_{\rm space}(t)\ket{2s-t+s_0}$, we have the constraint
that $|\vec{a}|=|\vec{a}'|$, i.e. the final position of the walker for the
braids $B_{\vec{a}}$, $B_{\vec{a}'}$ is the same, and the trace over the
coin DOF is non zero only if $a_t=a'_t$, i.e. the final step
is in the same direction for both paths.  The result is:
\begin{equation}
p(s,t)=\frac{1}{d^{(n-1)}2^t}\sum_{\{\vec{a},\vec{a}';\ |\vec{a}|=|\vec{a}'|=s,
a_t=a'_t\}}(-1)^{z(\vec{a},\vec{a}')} \langle L \rangle
 \end{equation}
%The links $L= \langle ({B_{\vec{a}'}^t}^\dagger
% B^t_{\vec{a}})^\text{Markov}\rangle$ are closures of the so called pure braid group ${\mathcal K}_n$ since
% the components begin and end at the same location.
%  
%\section{Kauffman brackets and Jones polynomials for $SU(2)_2$}
%\label{asymp}
which for $d=1$ and $\langle L \rangle=1$ reduces to the usual quantum walk distribution. 
\begin{figure}[h]
\begin{center}
\includegraphics[width=\columnwidth]{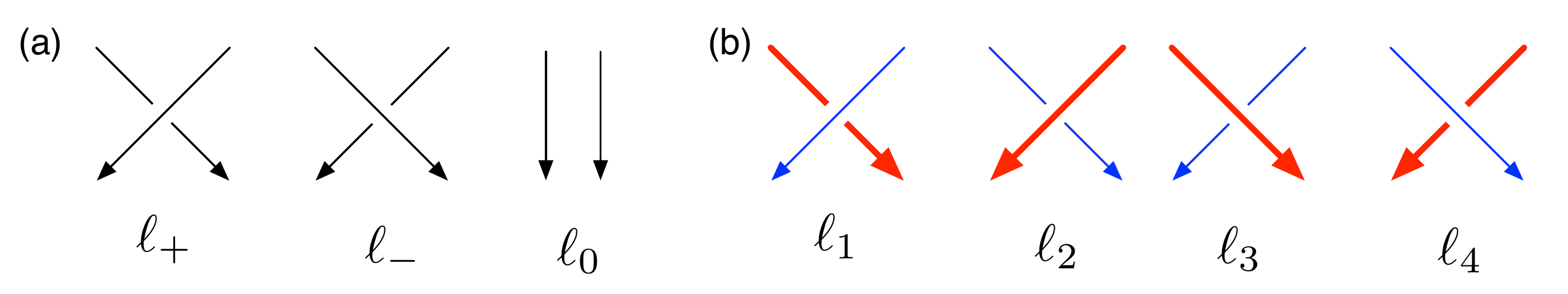}
\caption{\label{fig:2}(a) The writhe of a link is the difference of the number of positive crossing and negative crossings: $w(L)=\#\ell_+-\#\ell_-$.  (b) The linking number of two components, shown here as thick and thin components, is determined by the relative number of four types of crossings $lk(L_{thin},L_{thick})=\frac{\#\ell_1+\#\ell_2-\#\ell_3-\#\ell_4}{2}=\#\ell_1-\#\ell_4=\#\ell_2-\#\ell_3$.}
\end{center}
\end{figure}

The problem of computing the distribution thus reduces to computing statistics 
of a quantum link invariant.  
The Kauffman bracket of a link $\langle L \rangle(A)$ is a Laurent
polynomial in the argument $A$ that is an invariant for framed, unoriented
links and is framing dependent.   It can be related to the Jones polynomial
$V_L$, which is an invariant for framed, oriented links but
which is framing independent.  The relation is achieved by introducing an
orientation to each component in $L$ and suitably normalizing:
\[
\langle L \rangle (A)|_{A\rightarrow q^{-1/4}}=(-q^{3/4})^{w(L)}V_L(q)
\]
in such a way that the framing dependence of the Kauffman bracket is exactly canceled by the multiplicative factor involving the writhe $w(L)$ of $L$
(defined in Fig.\ref{fig:2}a). The Jones
polynomial of a link $L$ with variable $q=e^{i2\pi/(k+2)}$  was shown by Witten \cite{Witten} to be equal to
the expectation value of the product of the path ordered Wilson loops along
the components of the links of $L$ in $SU(2)_k$ Chern-Simons theory.

At the special value $q=i(k=2)$, where the links represent
braiding of anyons in the Ising model,
 the Jones polynomial of a link can be
related to a simpler knot invariant known as the $\mathtt{arf}$ invariant
\cite{Murakami}.  Specifically,
\[
V_L(i)=\left\{\begin{array}{c} \sqrt{2}^{(\#(L)-1)}(-1)^{\mathtt{arf}(L)}
\quad
\mathrm{if}\  L\ \mathrm{proper} \\\quad\quad\quad0\quad\quad\quad\quad\quad \quad\mathrm{if}\  L\ \mathrm{not\
proper}\end{array}\right.
\]
where the number of components $\#(L)=n$ here. An oriented link is
proper if each component $L_k$ evenly links the union of other components,
i.e. $\sum_{j\neq k}lk(L_j,L_k)=0\bmod 2 \forall j$.  The linking number of
two components is defined in Fig. \ref{fig:2}b and can be computed in
polynomial time.  The advantage of this expression is that
$\mathtt{arf}(L)\in\{0,1\}$ of a link can be computed in polynomial time in
the crossing number of a braid presentation. Hence, unlike the generic case, the Jones polynomials at value
$q=i$ can be evaluated in polynomial time  \cite{Jaeger}.  However, 
the number of links contributing to the weight $p(s,t)$ is $\binom{t-1}{s}^2+\binom{t-1}{s-1}^2$ which is exponential in $t$
so an efficient computation is not a priori available.  

%The $\text{arf}$ of a knot $K$, which is a single component link, is equal
%to zero if $K$ is pass equivalent to the unknot and is equal to one if $K$
%is pass equivalent to the trefoil knot {\bf ADD REF}.  It can be related the
%the Alexander polynomial for knots $\Delta_K$:
%\[
%\text{arf}(K)=\frac{(\Delta_K(-1))^2-1}{8}\bmod 2
%\]
%which can be computed in time polynomial in the crossing number of a braid
%presentation of the knot \cite{Alexander}.     For multicomponent links $L$,
%the $\text{arf}$ has a more complicated definition in terms of a knot that
%is related to the link.

In the special case the link $L$ has all pairwise linking numbers even, i.e. it is totally proper, then  there is a three local formula for the $\mathtt{arf}$ invariant:
\begin{equation}
\mathtt{arf}(L)=\displaystyle{\sum_i c_1(L_i)+\sum_{i<j}c_2(L_i,L_j)+
\sum_{i<j<k}c_3(L_i,L_j,L_k)}\bmod 2
\label{simparf}
\end{equation}
where $c_s(\Gamma)$ is the coefficient of $z^{s+1}$ in the Alexander-Conway polynomial
of the $s$ component sublink $\Gamma$ \cite{Murakami, Hoste}.  

There is structure to the links in the anyonic quantum walk trajectories
that simplifies Eq. \ref{simparf}.
First, if $L$ is proper than it is
totally proper.  This is due to the following:  if we consider the set of all the sums
$S=\{\sum_{k\neq j}lk(L_j,L_k)\}_{j=1}^n$, this is equal to
$S=\{lk(L_j,L_w)\}_{j\neq w}\sqcup \{\sum_{k\neq w}lk(L_w,L_k)\}$ where $L_w$ is
the walker's component. The condition of $L$ being proper is that every
member of $S$ is an even integer.  Now, the total linking
number is $\sum_{i<j} lk(L_i,L_j))=\sum_{k\neq w}lk(L_w,L_k))=0$. This
quantity is zero since for the forward half of the braid, i.e.
$B^t_{\vec{a}}$, all contributions to links are positive (counterclockwise
braiding) and for the latter half, ${B^t_{\vec{a}'}}^{\dagger}$,  there is
the same number of clockwise braids all of whom contribute with a negative
sign, so $S=\{lk(L_j,L_w)\}_{j\neq w}\sqcup \{0\}$.   But the condition that $L$
is totally proper is precisely that every member of the set
$\{lk(L_j,L_w)\}_{j\neq w}$ is an even integer.  Hence if $L$ is proper
then it is totally proper. Second, the writhe of
any link $ L$ is zero since the total linking number is zero and the braids
act on components all with the same orientation (see Fig.\ref{fig:1}b). 

With regard to the terms in Eq. (\ref{simparf}), note that due to causality of the worldlines there is no self linking, i.e. $c_1=0$.
The pairwise contribution to the $\mathtt{arf}$ is also zero since the sum of $c_2(L_j,L_k)$ is 
even as we now show.  It is only necessary to
consider links involving the walker $L_w$ and every other component $L_j$
since the non-walker components are not directly linked.  Such a link
is the braid closure of a braid in the two component braid group
$\mathcal{B}_2$ with one generator $b$ so that link can be written, 
$(L_w,L_j)=(b^{m})^\text{Markov}$ where $m=2\times lk(L_w,L_j)$.   To
compute the two point invariant we use the defining Skein relation for the
Alexander-Conway polynomial
\begin{equation}
\nabla_{\ell_+}-\nabla_{\ell_-}=z\nabla_{\ell_0}
\end{equation}
with the notation defined in Fig. \ref{fig:2}a. Moreover, $\nabla_{O}=1$ which states that if one of the components is unlinked from the others then it can be removed with a multiplicative factor of $1$.  The polynomial for any link $L$ can be written as $\nabla_L(z)=\sum_{i=0}^{\infty}a_i z^i$.  For our pairwise component links $(L_w,L_j)$ one can solve for the polynomial by recursion:
\[
\nabla_{(L_w,L_j)}= f_{|m|}(z)+f_{|m|-1}(z),
\]
where
\[
\begin{array}{lll}
f_m(z)&=&\frac{1}{2^{m+1}\sqrt{z^2+4}}\Big((z-\sqrt{z^2+4}) \big((z+\sqrt{z^2+4})^{m}\\
&&-(z-\sqrt{z^2+4})^{m}\big)\Big).
\end{array}
\]
From the coefficient of the cubic term of $\nabla_{(L_w,L_j)}$ we obtain
\[
c_2(L_w,L_j)=lk(L_w,L_j)  (lk(L_w,L_j)^2-1)/6.
\]
Note that the linking numbers are always even,
$lk(L_w,L_j)=2n_j$ with $n_j$ an integer, so the sum can be written as
\[
\begin{array}{lll}
\sum_{j<k}c_2(L_j,L_k) &=&
\frac{8}{6} \sum_{j\neq w} n_j^3 - \frac{8}{6} \sum_{j\neq w} n_j \\
%&=& 8 \sum_{j\neq w} \frac{n_j (n_j^2 - 1)}{6}\\
&=& 8 \sum_{j\neq w} \binom{n_j+1}{3}\quad  \in 2\mathbb{N}.\\
\end{array}
\]
Here we used the facts that the sum of the linking numbers is zero,
so that $\sum_{j\neq w}n_j=0$, and the
binomial coefficient is always an integer.  Finally, the triple component invariant $c_3(L_r,L_s,L_t)$ is known as the Milnor invariant. 
It counts the number of Borromean rings in three component sublinks  \cite{Kirby} and has nontrivial contribution to the $\mathtt{arf}$ of our links.  

This analysis allows us to express the probability distribution in terms of simple properties of the anyonic walk:
\begin{equation} \label{eqn:probability}
p(s,t)=\sum_{
\begin{array}{c} {\scriptstyle \{\vec{a},\vec{a}'\in\{0,1\}^{\otimes t}; }
 \\ {\scriptstyle\ |\vec{a}|=|\vec{a}'|=s, a_t=a'_t \}} \end{array}}
%\{\vec{a},\vec{a}'\in\{0,1\}^{\otimes t};\\
%\ |\vec{a}|=|\vec{a}'|=s,
%a_t=a'_t\}}}
 \left\{
\begin{array}{c}\frac{(-1)^{z(\vec{a},\vec{a}')+\tau(\vec{a},\vec{a}')}}{2^t}\quad
L\ \mathrm{proper} \\
\quad \quad 0
\quad\quad\quad L\ \mathrm{not\
proper}\end{array}\right.
\end{equation}
where the link $L$ is defined in Eq. (\ref{modfL}) and the sum of
Milnor invariants over sublinks $(L_r,L_s,L_t)$ of $L$ is
$\tau(\vec{a},\vec{a}')=\sum_{r<s<t}c_3(L_r,L_s,L_t)$.  When all links are proper
and $\tau(\vec{a},\vec{a}')$ is even then $p(s,t)\rightarrow p_{\rm QW}(s,t)$ the quantum walk distribution and when all non-mirror paths (i.e. $\vec{a}\neq\vec{a}'$) are non proper then
$p(s,t)\rightarrow p_{\rm RW}(s,t)$ the
classical random walk distribution.
 
To probe the behaviour of $p(s,t)$ we initially perform exact numerical simulations.  Note
that $z(\vec{a},\vec{a}')$ is a sum of local characteristics of the links, while $\tau(\vec{a},\vec{a}')$ is a sum of non-local ones.  To compute quantities which involve pairs or triples of components, we make
use of the braid word representations of the links Eqs.
(\ref{braidworda},\ref{modfL}) determined by the coin histories
$(\vec{a},\vec{a}')$. The linking number $lk(L_w,L_j)=(\# b_j-\#b^{\dagger}_j)/2$
when the walker's component is to the right of $j$ ($w>j)$ and 
$lk(L_w,L_j)=(\# b_j-\#b^{\dagger}_j)/2$
otherwise.
%  is one half the For example, to compute of the linking number
%$lk(L_w,L_j)$ for $L$ we write down the braid word ${B_{\vec{a}'}^t}^\dagger
%B^t_{\vec{a}}$ and count instances of crossings of those two components:
%\[
%lk(L_w,L_j)=\left\{\begin{array}{c}\frac{\# b_j-\#b^{\dagger}_j}{2}\quad
%w>j\ ({\rm walker\ on\ right}) \\\frac{\#
%b_{j-1}-\#b_{j-1}^{\dagger}}{2}\quad w<j\ ({\rm walker\ on\
%left})\end{array}\right.
%\]
If $lk(L_w,L_j)$ is an odd integer for any component $L_j$ then
$L=({B_{\vec{a}'}^t}^\dagger B^t_{\vec{a}})^\text{Markov}$ is not proper.
\begin{figure}[h]
  \begin{center}
    \includegraphics[width=8.5cm]{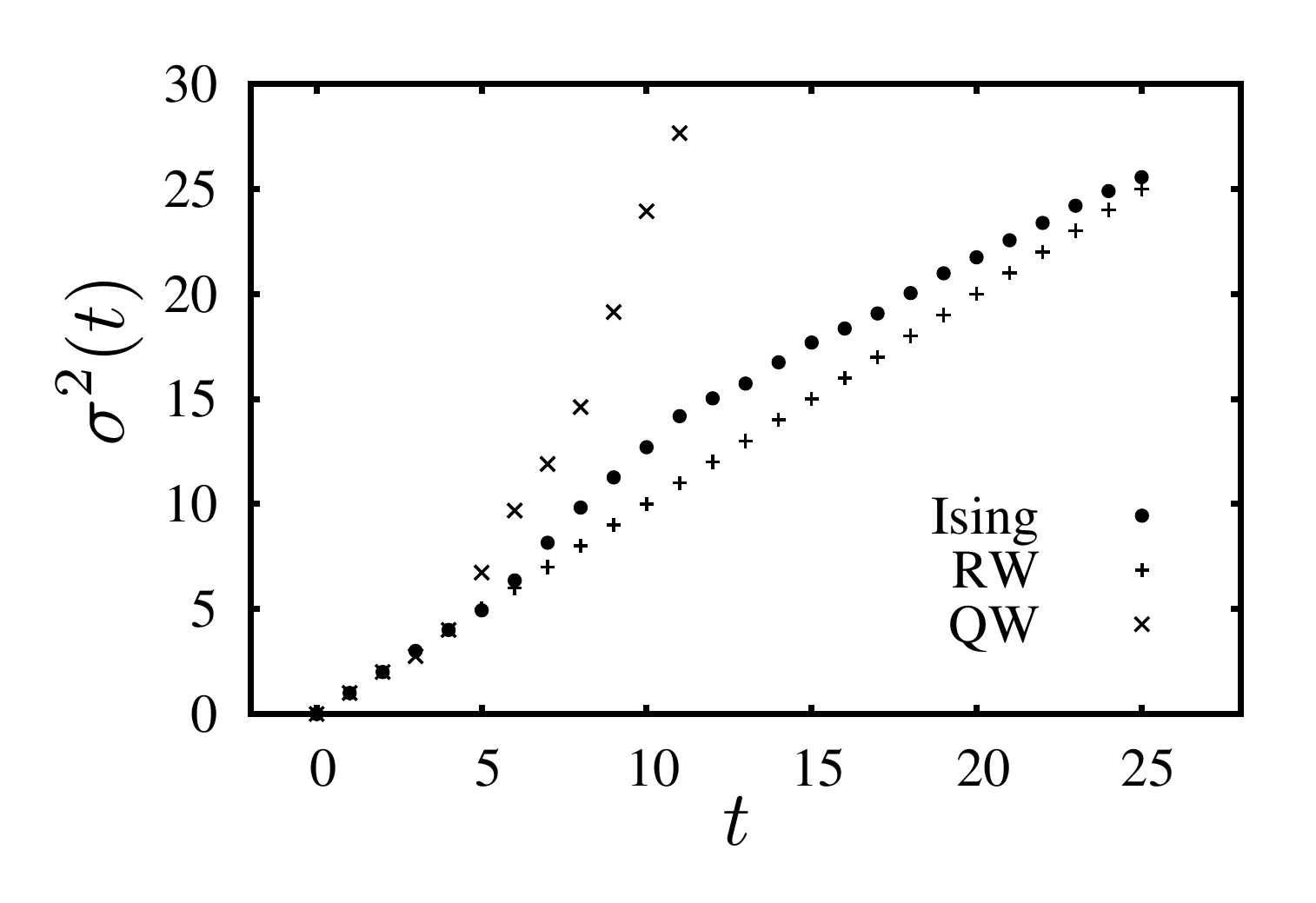}
    \caption{\label{fig:prob-variance}  The time evolution of the variance of the spatial distribution $p(s,t)$ (Eq. \ref{eqn:probability}) for the Ising anyonic walk and the corresponding classical and quantum walk evolutions with the same initial state.}
  \end{center}
\end{figure}
The Milnor invariant $c_3(L_r,L_w,L_s)$ where the components are ordered in
position with $r<w<s$, can be computed by writing down the braid word
$({B_{\vec{a}'}^t}^\dagger B^t_{\vec{a}})^\text{Markov}$ and deleting all
instances of braid word generators not in the set
$\{b_{r},b_{s-1},b_{r}^{\dagger},b_{s-1}^{\dagger}\}$.  The remaining braid
word $B(L_r,L_w,L_s)$ involves just two generators and their adjoints.  We
can think of this as a braid word on the three strands braid group. For
the spin-1/2 irrep of the Ising model, the representation of these
generators is:
\[
b_r=-e^{-i\pi/8}\left(\begin{array}{cc}1 & 0 \\0 & i \end{array}\right),\quad
b_{s-1}=-\frac{e^{i\pi/8}}{\sqrt{2}}
\left(\begin{array}{cc}1 & -i \\ -i & 1 \end{array}\right).
\]
Note that $b_r^2=e^{-i\pi/4}\sigma^z$ and $b_{s-1}^2=e^{-i\pi/4}\sigma^x$ where $\sigma^{x,z}$ are the Pauli
operators.  For proper links the product
of this representation of the braid generators in $B(L_r,L_w,L_s)$ will be
$\pm {\bf 1}$ with the sign carrying the value $(-1)^{c_3(L_r,L_w,L_s)}$.  The Milnor invariant 
for the other two orderings 
of components is calculated analogously.
%For example, for the link in Fig. \ref{fig:1}b, $B(L_r,L_w,L_s)=b_r^{\dagger\
%2}b_{s-1}^{\dagger\ 3}b_{s-1}b_r^2b_{s-1}^2=\sigma^z\sigma^x\sigma^z\sigma^x=-{\bf 1}$ corresponding to
%a single Borromean ring. For $c_3(L_w,L_r,L_s)$ (where the
%components are ordered in position with $w<r<s$) then before the first
%instance of $b_{r},b_{s-1}$ in the braid word we will have an instance of
%$b_{r-1}$, similarly, after the last instance
%$b_r^{\dagger},b_{s-1}^{\dagger}$ we will have an instance of
%$b_{r-1}^{\dagger}$.  But the computed quantity above $\pm {\bf 1}$ is
%invariant under conjugation.  A similar argument using conjugation by $b_s$
%applies to the case where the components are ordered in position with
%$r<s<w$.
%   In Ref. \cite{Mellor} a similar observation involving looking for
%clusters of the group commutator of the square of two braid word generators
%(or their inverses) is made in order to relate Milnor triple points to
%singular links in Borromean clasp theory.
We have calculated the distribution for walks up to $t = 25$.  The variance
$\sigma^2(t) =\langle s^2 \rangle - \langle s \rangle^2$ where the expectation value is
$\langle O(s) \rangle\equiv\sum_{s}O(s)p(s,t)$, is plotted in Fig. \ref{fig:prob-variance} and it
quickly approaches the linear random walk variance.  Using the total variation distance between two distributions $p(s,t)$ and $f(s,t)$ defined $\Delta(p,f)\equiv\frac{1}{2}\sum_s |p(s,t)-f(s,t)|$, at $t=25$ we find $\Delta(p,p_{\rm QW})=0.34$ while $\Delta(p,p_{\rm RW})=0.04$.

We now show that asymptotically the Ising anyonic walk behaves classically.
The essential reason is the rapidly decreasing density of proper links in the regime where
the quantum walk distribution has dominant support.
%First we rewrite Eq. \ref{eqn:probability} as
%\begin{equation}
%p(s,t)_{\rm AQW}=\frac{1}{2^{t}}\Bigg(\displaystyle{\sum_{\vec{a}=\vec{a}'}}1+
%\displaystyle{\sum_{\{\vec{a}\neq\vec{a}';\ |\vec{a}|=|\vec{a}'|=s,
%a_t=a'_t\}}(-1)^{z(\vec{a},\vec{a}')+arf(L)}} \Bigg)\\
%\end{equation}
% In order to upper bound the variance we can assume that the arfs of all the proper links 
% are even.
% It was shown in {\bf [need ref]} that the quantum walk with the single coin has
% the largest scaling in variance possible for a linear quantum mechanical walk evolution (i.e. no
% other DOF attached to the system can make it faster) so making the arf even can only make the estimate of the variance of the anyonic walker larger.
In order to upper bound the variance we can assume that $\tau(\vec{a},\vec{a}')$ of all the proper links are even and that there is no correlation between being proper and $z(\vec{a},\vec{a}')$ for non mirror paths.  Calling the resulting distribution $\tilde{p}(s,t)$ we have
\begin{equation}
\tilde{p}(s,t)=p(s,t)_{\rm RW}+p_{\rm prop}(s,t) [p(s,t)_{\rm QW}-p(s,t)_{\rm RW})],
\label{boundprob}
\end{equation}
where $p_{\rm prop}(s,t)$ is the density of proper links for non-mirror paths ($\vec{a}\neq \vec{a}'$).
Since the walker's speed is constant, the maximum possible variance is quadratic, achieved up to a constant less than one by the quantum walk \cite{Nayak}, so this choice of distribution can only make the estimate of the variance of the anyonic walker larger, i.e. 
$\tilde{\sigma}^2(t)\geq \sigma^2(t)$.

In Ref. \cite{Nayak} it was shown that the distribution of the quantum walk with the same Hadamard coin flip and initial state as occurs here can be very well approximated asymptotically by the function $p_{\rm QW}'(\alpha,t)d\alpha=\frac{(1-\alpha)d\alpha}{\pi(1-\alpha^2)\sqrt{1-2\alpha^2}}$ where  $\alpha=(s_0-s)/t$ and is restricted to the interval $[-\frac{1}{\sqrt{2}},\frac{1}{\sqrt{2}}]$.  Outside this interval the distribution falls off exponentially with $t$ as does $p_{\rm QW}(s,t)$.  Restricting to this interval and using the fact that a proper link must have all linking numbers even, it is shown in the Appendix  that the density of non-mirror proper links is $p_{\rm prop}(s,t)<C/t^2$ for some constant $C$ independent of $s,t$.  The position moments with respect to $\tilde{p}(s,t)$ are
\[
\begin{array}{lll}
\langle (s-s_0)\rangle&\approx&\scriptstyle{\int_{1/\sqrt{2}}^{-1/\sqrt{2}}} p_{\rm prop}(s,t)t\alpha p_{\rm QW}'(\alpha,t)d\alpha\\
&<& C(1-1/\sqrt{2})/t,\\
\langle (s-s_0)^2\rangle&\approx& t(1-C/t^2)+\scriptstyle{ \int_{1/\sqrt{2}}^{-1/\sqrt{2}}} p_{\rm prop}(s,t)t^2\alpha^2p_{\rm QW}'(\alpha,t)d\alpha\\
&<& t(1-C/t^2)+(1-1/\sqrt{2})C,
\end{array} 
\]
Thus $\tilde{\sigma}^2(t)< t +O(1)$.

  Similarly, to obtain a lower bound for the dispersion we can assume a distribution of the form of Eq. \ref{boundprob} but chose a probability distribution $f(s,t)$ with minimum variance to replace $p_{\rm QW}(s,t)$.  Picking $f(s,t)=\delta_{s,s_0}$ (zero variance) and calling the resulting distribution $\tilde{\tilde{p}}(s,t)$ then $\tilde{\tilde{\sigma}}(t)^2> t(1-C/t^2)$.   By the inequalities $\tilde{\tilde{\sigma}}^2(t)\leq\sigma^2(t)\leq \tilde{\sigma}^2(t)$ we find 
\begin{equation}
\lim_{t\rightarrow \infty}\frac{\sigma^2(t)}{t}=1,
\end{equation}
i.e. asymptotically the Ising anyonic walk has linear dispersion with coefficient 1.

In conclusion, we have studied the dynamical behavior of a mobile non-Abelian anyon which becomes entangled with its environment purely by statistical interactions.  We find that for the case of Ising anyons the decoherence is strong enough to completely wash out the quantum mechanical interferences and reduce the dynamics to a classical stochastic process. 
   It would be of interest to extend this analysis to anyons which are spin-$1/2$ irreps for $SU(2)_{k>2}$ models since it is known that for $k>2,k\neq4$, the braiding evolutions densely span the fusion space while for $k=2$ (Ising anyons) and $k=4$ they do not \cite{FLW}.  It has been shown that for a $t$ step quantum walk subject to decoherence in its position at a rate $p_{\rm meas}>C'/t$ for some constant $C'$, the evolution approaches classical behavior \cite{KendonII}.  We might then conjecture that since the braiding generators that entangle new fusion degrees of freedom can be translated into a measurement error $p_{\rm meas}\sim1/k^2$ (for $k\gg 1$), then for  $t>k^2$ the walk would behave classically.  In the limit $k\rightarrow \infty$ the particles are fermions and we recover quantum walk behavior with quadratic dispersion.
In general a mobile anyons experiences an environmental coupling that is highly non-local and non-Markovian and its behavior sheds light on the statistical dynamics of more complex systems. 

{\it Acknowledgments---}
LL and VZ thank the EU grant EMALI for support during their visit at Leeds University. JKP would like to thank the Royal Society.  GKB received support from the Australian Research Council and 
 from the European Community's Seventh Framework Programme (FP7/2007-2013) under grant agreement n¡ 247687 (Integrating Project AQUTE).

\appendix
\label{app}

\section{Appendix:  Upper bound for $p_{\rm prop}(s,t)$}

We will show that $p_{\rm prop}(s_0,t)$, the probability of a randomly chosen link $L$ to be proper (mirror paths excluded), is bounded from above by a function $\frac{C}{t^2}$ under certain assumptions. We first consider the number of links with even linking numbers and derive $p_{\rm prop}(s_0,t)$ as the proportion of these links with respect to all links. The result holds for paths which end up on the initial site but numerical results suggest that
$p_{\rm prop}(s,t) < p_{\rm prop}(s_0,t)$ if $s \neq s_0$, provided $s$ is not to close to the boundaries at $s_0\pm t$.  In particular, it is true inside the domain $\frac{s_0-s}{t}=[-\frac{1}{\sqrt{2}},\frac{1}{\sqrt{2}}]$.

\begin{figure}[h]
\begin{center}
\includegraphics[width=\columnwidth]{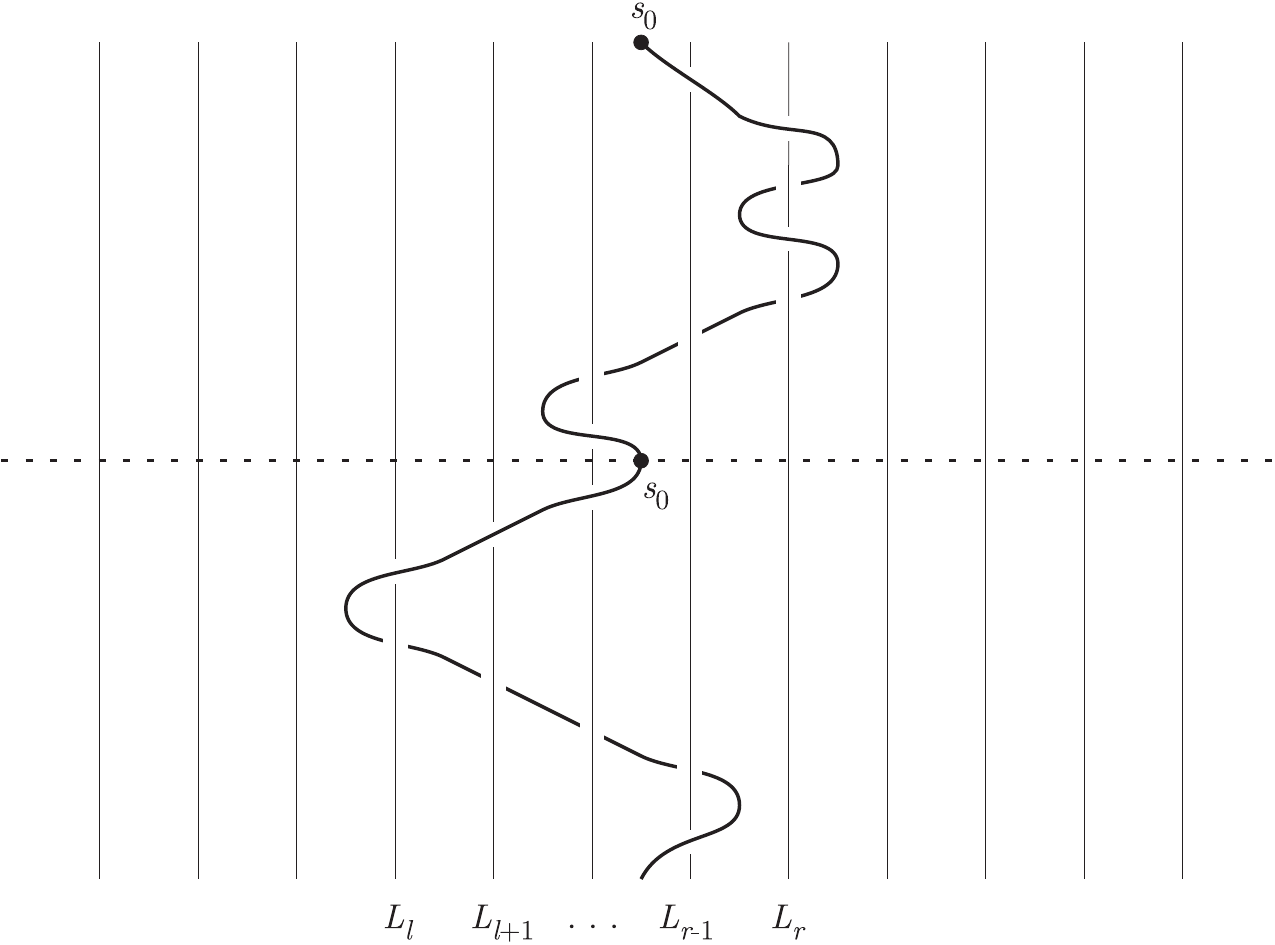}
\caption{\label{fig:path Ll Lr} A generic path reaching after $t$ steps (here $t=8$) the position $s_0$. The path braids with all the link components $L_l,...,L_r$. We say that its width $w$ is $r-l+1$ (here $w=5$).}
\end{center}
\end{figure}

Suppose a path $(\vec{a},\vec{a}')$ leading after $t$ steps to the position $s_0$ links with link components $L_j$, $j\in \{l,...,r\}$ (Fig.\ref{fig:path Ll Lr}). 
For the corresponding link $L$ to be proper, all the linking numbers $lk(L_w,L_l),..., lk(L_w,L_r)$ must be even. Let $p_{\rm e}(t,j)$ be the probability that $lk(L_w,L_j)$ is even. We assume that for large enough (fixed) $t$, the probabilities $p_{\rm e}(t,j)$ can be treated as independent.   This is justified inside the interval $\frac{s_0-s}{t}=[-\frac{1}{\sqrt{2}},\frac{1}{\sqrt{2}}]$, where the number of paths that contribute to the anyonic walk density at each point $s$ is exponential in $t$.  The linking number of the walker with any particular component can change even/odd parity by a simple deformation and within the typical width (i.e. number of components touched by the walker) of the path there are an exponential number of such deformations hence the linking numbers of the walker with those components are well approximated as independent quantities.  

Then we can write the probability for the link $L$ to be proper as
\begin{equation} \label{}
p_{\rm prop}(s_0,t) = p_{\rm e}(t,l)\cdot ...\cdot p_{\rm e}(t,r) ~.
\end{equation}
If we denote
\begin{equation} \label{}
\rho \equiv \max_j p_{\rm e}(t,j) ~,
\end{equation}
where $j$ runs over all link components $L_j$ that braid with some $t$ step long path leading to $s_0$, we can estimate from above
\begin{equation} \label{eqn:proper link probability}
p_{\rm prop}(s_0,t) \leq \rho^w ~,
\end{equation}
where $w=r-l+1$ we will call the \emph{width} of the link.

\begin{figure}[h]
\begin{center}
\includegraphics[width=\columnwidth]{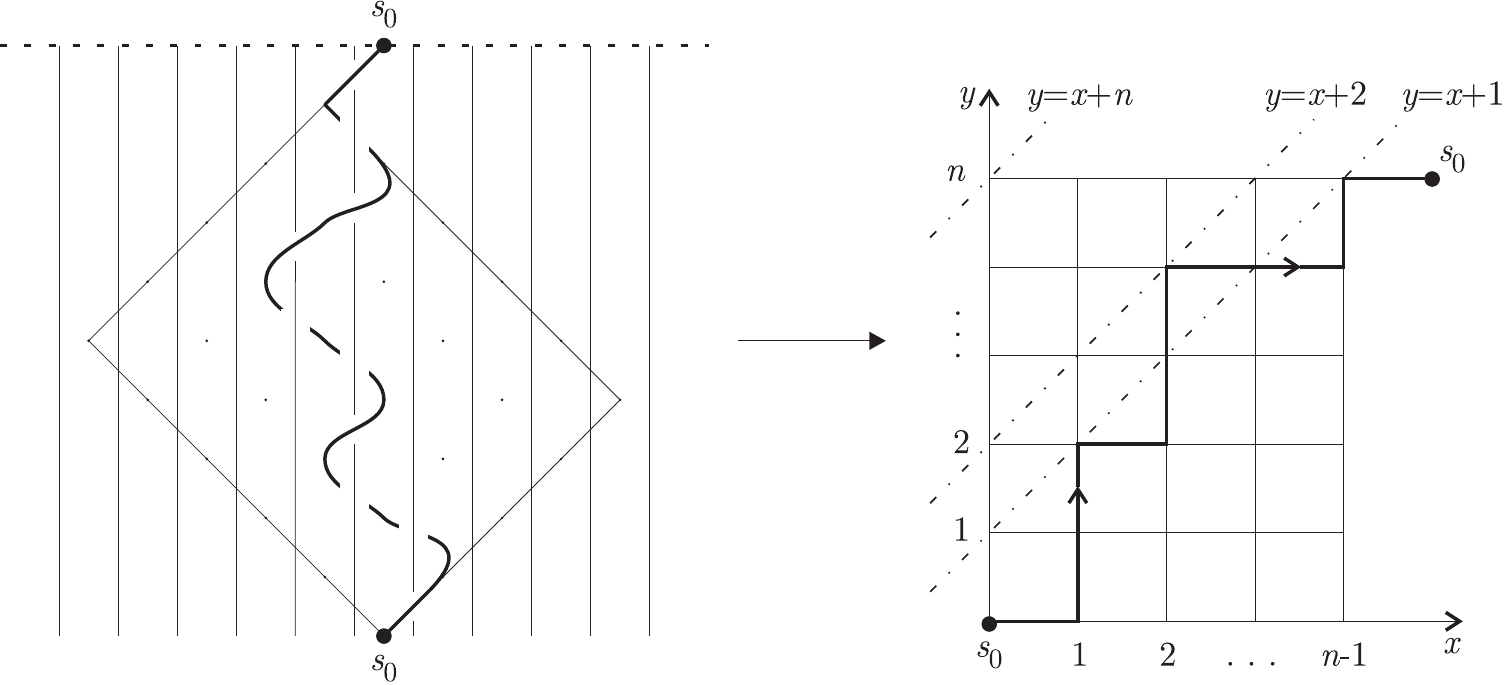}
\caption{\label{fig:path on a lattice} Any half-path (i.e. either "bra" or "ket" history of the walker) can be thought of as being realized on a two dimensional lattice, where only the steps $\uparrow$ and $\rightarrow$ are allowed. If we fix the number of steps $t$ (presume $t$ even), the position reached after $t$ steps $s=s_0$ and the last coin outcome $a_t=1$, we can draw the possible half-paths on a lattice $n-1$ by $n$, where $n=\frac{t}{2}$. The lattice paths that touch the diagonal $y=x+w$ ($1\leq w\leq n$), but don't touch $y=x+w+1$, braid with at least $w$ link components. "At least", because they certainly braid with the $w$ components to the left of $s_0$, but they might be (and typically are) spread to the right of $s_0$ as well.}
\end{center}
\end{figure}

The paths $(\vec{a},\vec{a}')$ relevant for a walk of $t$ steps must satisfy $a_t=a'_t$. Let us consider the paths leading to $s_0$ (we presume $t$ is even), for which $a_t=a'_t=1$. It is useful to depict $\vec{a}$ as a lattice path on a $n-1 \times n$ lattice ($n=\frac{t}{2}$) with allowed steps $\uparrow$ and $\rightarrow$ (Fig.\ref{fig:path on a lattice}). 
The number of all such paths is
\begin{equation} \label{eqn:number of all lattice paths}
\binom{2n-1}{n-1} \equiv \#(all) ~.
\end{equation}
The number of lattice paths that touch the diagonal $y=x+w, w \in \{1,...,n\}$ is in our case of $n-1$ by $n$ lattice
\begin{equation} \label{}
\binom{2n-1}{n-w} \equiv \#(w)~.
\end{equation}
To derive this result we notice that $\#(1)=\#(all)$ and for $w \in \{2,...,n\}$ we use the \emph{Andr\'{e} reflection principle}, described, for example, in \cite{Mohanty}.

Consider the set $\mathcal{P}_w$ of lattice paths that touch the diagonal $y=x+w$ but do not touch the further diagonal $y=x+w+1$. The sets $\mathcal{P}_w$ are for distinct $w \in \{1,...,n\}$ obviously disjoint. The number of paths in $\mathcal{P}_w$ is 
\begin{equation} \label{eqn:number of paths in Pw}
|\mathcal{P}_w| = \#(w)-\#(w+1) = \binom{2n-1}{n-w}-\binom{2n-1}{n-w-1} ~.
\end{equation}
We use the convention that $\binom{m}{k}=0$ if $k<0$.  Furthermore, the union $\bigcup_{w=1}^{n} \mathcal{P}_w$ comprises all the paths on $n-1$ by $n$ lattice, which follows from 
\begin{equation}
\sum_{w=1}^{n} |\mathcal{P}_w| = \sum_{w=1}^{n} \left( \binom{2n-1}{n-w}-\binom{2n-1}{n-w-1} \right) = \binom{2n-1}{n-1} = \#(all) ~.
\end{equation}

\begin{figure}[h]
\begin{center}
\includegraphics[width=\columnwidth]{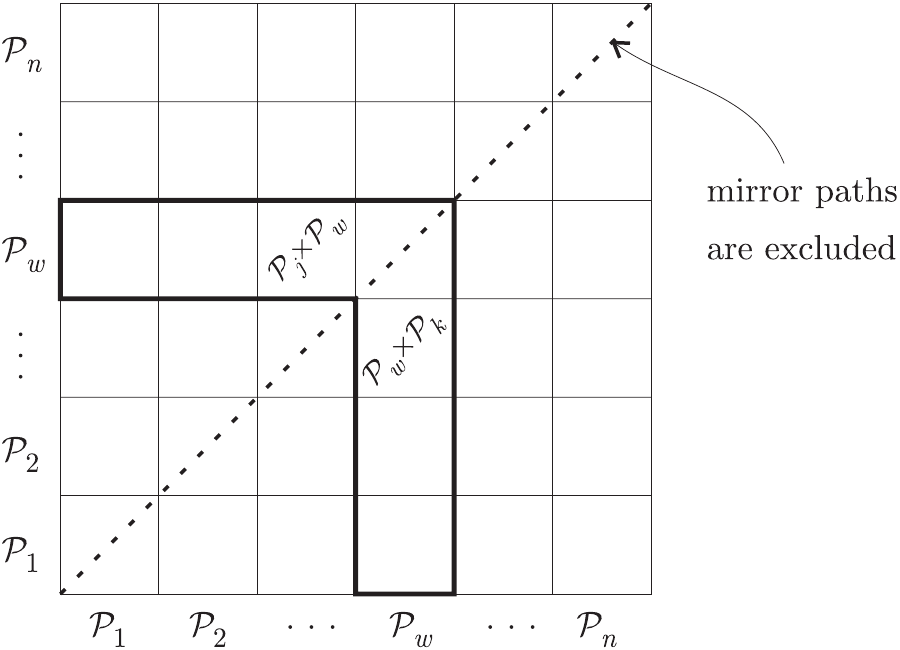}
\caption{\label{fig:chessboard} The set of paths that after $t$ steps reach the position $s=s_0$, with the last coin outcome $a_t=1$, can be expressed in terms of the sets $\mathcal{P}_{w}$ as $\left[\bigcup_{w,w'} \mathcal{P}_{w} \times \mathcal{P}_{w'}\right] \backslash M = \left[ \bigcup_{w=1}^n \left(\mathcal{P}_{w} \times \mathcal{P}_{w}\right) \cup \left( \bigcup_{j=1}^{w-1} \mathcal{P}_{j} \times \mathcal{P}_{w}\right) \cup \left( \bigcup_{k=1}^{w-1} \mathcal{P}_{w} \times \mathcal{P}_{k}\right) \right] \backslash M$, where $M \equiv \{(\vec{a},\vec{a}') | \vec{a}=\vec{a}'\}$.}
\end{center}
\end{figure}

A path $(\vec{a},\vec{a}')$ consists of two lattice paths, $\vec{a}$ and $\vec{a}'$, where $\vec{a} \in \mathcal{P}_w$ and $\vec{a}'  \in \mathcal{P}_{w'}$ for some $w,w' \in \{1,...,n\}$, i.e. $(\vec{a},\vec{a}') \in \mathcal{P}_w \times \mathcal{P}_{w'}$. Let's realize that the width of the path $(\vec{a},\vec{a}')$ is at least $\max\{w,w'\}$ (see again Fig.\ref{fig:path Ll Lr} and Fig.\ref{fig:path on a lattice}). The probability that the path $(\vec{a},\vec{a}')$ is proper can thus be estimated from above (using (\ref{eqn:proper link probability})) as
\begin{equation} \label{eqn:proper link probability bound}
p_{\rm prop}(s_0,t) \leq \rho^{\max\{w,w'\}} ~.
\end{equation}
Finally, the set of all non-mirror paths with the last step "1" reaching after $t$ steps the starting position $s_0$ can be expressed in terms of $\mathcal{P}_{w}$'s as shown in Fig.\ref{fig:chessboard}.

Now we can write an upper bound for $p_{\rm prop}(s_0,t)$ as the number of links with even linking
numbers divided by the number of all links:
\begin{equation} \label{eqn:proper vs all upper bound 1}
% \frac{N'_p(s_0,t)}{N'(s_0,t)} \leq
p_{\rm prop}(s_0,t) \leq
\frac{2 \left( \sum_{w=1}^n \rho^w \left(|\mathcal{P}_w|^2-|\mathcal{P}_w| + 2 \sum_{j=1}^{w-1} |\mathcal{P}_w| |\mathcal{P}_j| \right) \right)}{2 \big(\#(all)^2-\#(all)\big)} ~.
\end{equation}
The number 2 in both the nominator and the denominator accounts for the fact that we count both types of paths -- with $a_t=a'_t=0$ and $a_t=a'_t=1$ -- and use the symmetry between the two situations. We used the relation (\ref{eqn:proper link probability bound}) and the diagram at Fig.\ref{fig:chessboard}. Also, we didn't forget to exclude the mirror paths.

Before substituting to (\ref{eqn:proper vs all upper bound 1}), we rewrite (\ref{eqn:number of paths in Pw}) as
\begin{equation}\label{number of paths in Pw 2}
|\mathcal{P}_w| = \frac{(2n-1)!}{(n-w-1)!(n+w-1)!} \left( \frac{1}{n-w} - \frac{1}{n+w} \right) = \frac{w}{n} \binom{2n}{n-w}
\end{equation}
and (\ref{eqn:number of all lattice paths}) as
\begin{equation}\label{eqn:number of all lattice paths 2}
\#(all) = \frac{(2n-1)!}{n!(n-1)!} = \frac{1}{2} \binom{2n}{n} ~.
\end{equation}
Now (\ref{eqn:proper vs all upper bound 1}) reads
\[
\begin{array}{l}
% \frac{N'_p(s_0,t)}{N'(s_0,t)} \leq
p_{\rm prop}(s_0,t) \leq
\frac{\frac{1}{2} \binom{2n}{n}}{\frac{1}{2} \binom{2n}{n} - 1}
\frac{\sum_{w=1}^n \rho^w \left(\binom{2n}{n-w}^2-|\mathcal{P}_w| + 2 \sum_{j=1}^{w-1} |\mathcal{P}_w| |\mathcal{P}_j| \right)}{\frac{1}{2} \binom{2n}{n} \frac{1}{2} \binom{2n}{n}} ~,
\end{array}
\]
which we estimate from above using the relation between binomial coefficients
$\binom{2n}{n-w} \leq \binom{2n}{n}$ to get:
\begin{eqnarray}
% \frac{N'_p(s_0,t)}{N'(s_0,t)} &\leq&
p_{\rm prop}(s_0,t) &\leq&
\frac{\frac{1}{2} \binom{2n}{n}}{\frac{1}{2} \binom{2n}{n} - 1} \frac{\sum_{w=1}^n \rho^w \left(\frac{w^2}{n^2} + 2 \sum_{j=1}^{w-1} \frac{w}{n} \frac{j}{n} \right)}{\frac{1}{2} \frac{1}{2}} \nonumber\\
&=& \left( 1 + \frac{1}{\frac{1}{2} \binom{2n}{n} - 1} \right) \frac{4}{n^2} \sum_{w=1}^n \rho^w \left(w^2 + 2 w \sum_{j=1}^{w-1} j \right) \nonumber\\
&=& \left( 1 + \frac{1}{\frac{1}{2} \binom{2n}{n} - 1} \right) \frac{4}{n^2} \sum_{w=1}^n \rho^w w^3 ~,
\end{eqnarray}
where we used the formula $\sum_{j=1}^{w-1} j = \frac{1}{2} w (w-1)$. For $n \geq 2$ (i.e. $t \geq 4$)
\begin{equation}
\frac{1}{\frac{1}{2} \binom{2n}{n} - 1} \leq \frac{1}{\frac{1}{2} \binom{4}{2} - 1} = \frac{1}{5} ~.
\end{equation}
Thus
\begin{equation} \label{eqn:proper vs all upper bound 2}
% \frac{N'_p(s_0,t)}{N'(s_0,t)} \leq
p_{\rm prop}(s_0,t) \leq
\frac{6}{5} \frac{4}{n^2} \sum_{w=1}^n \rho^w  w^3 \leq \frac{6}{5} \frac{4}{n^2} \sum_{w=1}^\infty \rho^w  w^3 ~.
\end{equation}
The infinite sum on the right hand side converges for all $\rho \in [0,1)$ and one can find that
\begin{equation}
\sum_{w=1}^\infty \rho^w  w^3 = \frac{\rho(1+4\rho+\rho^2)}{(1-\rho)^4} ~.
% ~ \rho \in [0,1) ~.
\end{equation}
% Hence (\ref{eqn:proper vs all upper bound 2}) reads (recall that $n=\frac{t}{2}$)
% \begin{equation}
% \frac{N'_p(s_0,t)}{N'(s_0,t)} \leq
% \frac{1}{t^2} \frac{6\cdot 4\cdot 4}{5} \frac{\rho(1+4\rho+\rho^2)}{(1-\rho)^4} ~,
% \end{equation}
% where $\rho$, in general, is a function of time, $\rho=\rho(t)$. If, however, we assume (plausibly enough) that there exists $p_{max}$ such that $p(t) \leq p_{max} < 1$ for all $t$, we can finally write the desired upper bound
% \begin{equation} \label{eqn:proper vs all upper bound final}
% \frac{N'_p(s_0,t)}{N'(s_0,t)} \leq
% \frac{1}{t^2} \frac{96}{5} \frac{\rho(1+4\rho+\rho^2)}{(1-\rho)^4} = \frac{C}{t^2} ~,
% \end{equation}
% where $C$ doesn't depend on time (number of steps) $t$.
Hence we obtain the following upper bound (recall $n=t/2$):
\begin{eqnarray}
p_{\rm prop}(s_0,t) &\leq&
\frac{1}{t^2} \frac{6\cdot 4\cdot 4}{5} \frac{\rho(1+4\rho+\rho^2)}{(1-\rho)^4}\\
% &\leq& \frac{1}{t^2} \frac{96}{5} \frac{\rho(1+4\rho+\rho^2)}{(1-\rho)^4}\\
&=& \frac{C}{t^2} ~,
\label{eqn:proper vs all upper bound final}
\end{eqnarray}
where $C$ doesn't depend on time (number of steps) $t$.

\end{document}